\documentclass[namedreferences]{SolarPhysics}
\usepackage[optionalrh]{spr-sola-addons} 
\usepackage{url}             

\usepackage{graphicx}

\newcommand{\etal}{\textit{et al.}}
\newcommand{\eg}{\textit{e.g.}}

\newcommand{\be}{\begin{equation}}
\newcommand{\ee}{\end{equation}}
\newcommand{\beq}{\begin{eqnarray}}
\newcommand{\eeq}{\end{eqnarray}}

\newcommand{\aeta}[3]{  #1, {\it Astron. Astrophys.} {\bf  #2}, #3.}

\newcommand{\sph}[3]{   #1, {\it Solar Phys.} {\bf  #2}, #3.}
\newcommand{\sphp}[1]{  #1, {\it Solar Phys.}, in press.}

\newcommand{\arcsec}{^{\prime \prime}}

\begin{document}
\begin{article}
\begin{opening}

   \title{Do Long-Lived Features Really Exist in the Solar
   Photosphere?\\II. Contrast of Time-Averaged Granulation Images}

   \author{P.N. \surname{Brandt}$^{1}$ and
           A.V. \surname{Getling}$^{2}$}

   \runningauthor{Brandt and Getling}

   \runningtitle{Long-Lived Features in the Photosphere II}

   \institute{$^{1}$Kiepenheuer-Institut f\"ur Sonnenphysik,
              Sch\"oneckstrasse 6, 79104 Freiburg, Germany\\
              email: \url{pnb@kis.uni-freiburg.de}\\
              $^{2}$Institute of Nuclear Physics, Lomonosov
              Moscow State University, 119991 Moscow, Russia\\
              email: \url{A.Getling@mail.ru}\\
              }

   \date{Received 10 November 2007; accepted 31 January 2008}

\begin{abstract}
The decrease in the rms contrast of time-averaged images with the
averaging time is compared between four datasets: (1) a series of
solar granulation images recorded at La Palma in 1993; (2) a series
of artificial granulation images obtained in numerical simulations by
Rieutord \etal\ \cite{rieut}; (3) a similar series computed by
Steffen and his colleagues (see Wedemeyer \etal, \citeyear{stef}); (4) a random
field with some parameters typical of the granulation, constructed by
Rast \cite{rast}. In addition, (5) a sequence of images was obtained
from real granulation images using a temporal and spatial shuffling
procedure, and the contrast of the average of $n$ images from this
sequence as a function of $n$ is analysed. The series (1) of real
granulation images exhibits a considerably slower contrast decrease
than do both the series (3) of simulated granulation images and
the series (4) of random fields. Starting
from some relatively short averaging times $t$, the behaviour of the
contrast in series (3) and (4) resembles the $t^{-1/2}$ statistical
law, while the shuffled series (5) obeys the $n^{-1/2}$ law from
$n=2$ on. Series (2) demonstrates a peculiarly slow decline of
contrast, which could be attributed to particular properties of the
boundary conditions used in the simulations. Comparisons between the
analysed contrast-variation laws indicate quite definitely that the
brightness field of solar granulation contains a long-lived
component, which could be associated with locally persistent dark
intergranular holes and/or with the presence of quasi-regular
structures. The suggestion that the random field (4) successfully
reproduces the contrast-variation law for the real granulation (Rast,
\citeyear{rast}) can be dismissed.
\end{abstract}


\end{opening}

\section{Introduction}

As reported previously by a number of researchers, the brightness
field of solar granulation bears evidence for the presence of
features with lifetimes far exceeding those of granules.
In particular, Roudier \etal\ \cite{roudier} detected dark
``intergranular holes'' --- singularities in the network of
intergranular lanes, which could be observed for more than 45 minutes
and whose diameters range from 180 to 330 km. Similar features,
which persisted over up to 2.5 hours, were also observed by Hoekzema,
Brandt, and Rutten \cite{hbr} and Hoekzema and Brandt \cite{hb}.

A different form of long-term persistence, closely related to
spatial structuring, was revealed in granulation images averaged over
intervals of one to two hours. As noted by Getling and Brandt
\cite{gb1} and Getling (2006, Paper I), averaged brightness fields
consist of bright, nearly granular-sized blotches against a darker
background; they may form long-lived, quasi-regular systems of
concentric rings or parallel strips~--- ``ridges'' and ``trenches''
in the relief of the brightness field. Later, algorithmic
treatment showed that patterns of alternating bright and dark lanes
(varying in the degree of their regularity) are virtually ubiquitous
in averaged granulation images (Getling and Buchnev, \citeyear{gbu}).
Since the material upflows in the visible layers of the solar
atmosphere are brighter than the downflows, such patterns should
reflect the presence (and, possibly, widespread occurrence) of roll
convection in the upper convection zone. In particular, roll systems
could form the fine structure of larger-scale, supergranular and/or
mesogranular, convective flows.

Some other indications of a long-term spatial organisation in the
granulation field have also been revealed. Dialetis \etal\
\cite{dial} found that granules with longer lifetimes exhibit a
tendency to form mesogranular-scaled clusters. Muller, Roudier, and
Vigneau \cite{mul} also detected such clustering in the spatial
arrangement of large granules; they emphasised a plausible
relationship between the clusters and mesogranules. Roudier \etal\
\cite{roud5} reported their observations of a specific collective
behaviour of families (``trees'') of fragmenting granules. Such
families can persist for up to eight hours and appear to be related
to mesogranular flows. An imprint of the supergranular structure
can also be traced in the granulation pattern (Baudin, Molowny-Horas,
and Koutchmy, \citeyear{baudin}).

As a particular argument for the presence of long-lived features, we
noted (Getling and Brandt, \citeyear{gb1}) a conspicuously slow
decrease in the rms contrast of the averaged images with the
averaging time.

In his comment on our original paper (Getling and
Brandt,~\citeyear{gb1}), Rast~\cite{rast} suggested that the
quasi-regular structures detected in the granulation patterns are
merely of statistical nature and do not reflect the structure of real
flows. He artificially constructed a series of random fields with
some characteristic parameters typical of the solar granulation
(claiming nevertheless that this series ``is not an attempt to model
solar granulation'' and represents ``a completely random and changing
flow pattern'') and found that the law of contrast variation with the
averaging time does not change substantially if the real images are
replaced by such fields. On this basis, Rast \cite{rast} raised doubts on the
real presence of a long-term component in the granulation dynamics.

In view of attributing the features of regularity in the averaged
images to statistical effects, Rast \cite{rast} also put forward some other
arguments. Getling (Paper I) discussed them in detail and has
shown that the reasoning by Rast \cite{rast} does not contest the
possibility of a
physical origin of the quasi-regular structures. At the same time,
the particular issue of the contrast variation with the averaging
time was reserved for a special consideration.

This paper continues the publication of our study whose first part
was presented in Paper I and addresses precisely the contrast-variation
laws for various series of images.

Generally, the less variable the
pattern, the slower the contrast decline with the length of the
averaged sequence of images. Thus, the direct implication of a slow
decrease in the contrast of averaged images is the presence of
persistent features in the brightness field under study. The
contrast-variation law itself does not reflect the degree of spatial
order in this field; however, if the
regularity is associated with temporal persistence, the presence of
regular structures will flatten the contrast-variation curve. For
these reasons, we extend here the conceivable range of interpretation
of our results compared to that discussed in Paper I. Accordingly,
we have slightly modified the main title of the paper (``Do
quasi-regular structures really exist in the solar
photosphere?''), which was originally intended to be
common for both parts.

We compare here the laws of contrast variation with the averaging
time for real granulation images, images obtained by numerical
simulation of convection on a granular scale, the artificial fields
constructed by Rast \cite{rast}, and a series of ``shuffled'' images (obtained as
subfields of the original series of granulation images taken at
random positions in both spatial coordinates and at random times).
Our results confirm the presence of a long-lived component in the
granulation field, while the counterargument by Rast \cite{rast}
based on the contrast variation can be dismissed.

\section{The Data}

By the real granulation images, we mean the images of the La~Palma
series obtained by Brandt, Scharmer, and Simon (see Simon \etal,
\citeyear{simon}) on 5 June 1993 using the Swedish Vacuum Solar
Telescope (La Palma, Canary Islands). We analysed a seven-hour subset
of this series; the frame cadence was 21.03~s, and a
$43.5\times 43.5$~Mm$^2$ subfield ($480\times 480$ pixels with a pixel
size of 0.125$\arcsec$, or 90.6~km) was cut out of the original images.
In Paper I, we already
described some details of the technique of data acquisition and
pre-processing employed by Simon \etal\ \cite{simon}.

The random fields constructed by Rast~\cite{rast} are additive
superpositions of $192^2$ randomly disposed two-dimensional Gaussians
that vary randomly in amplitude and radius around given mean values.
These values and the time scale of the evolution were chosen so as to
mimic the corresponding parameters of the solar granulation. The
sequence of such fields represents a continuous time evolution with
the persistent emergence of ``new'' Gaussian peaks and disappearance
of ``old'' ones. The resulting images contain $192\times 192$ pixels.
In the language of the numerical correspondence between the
parameters of the random field and those of the granulation patterns,
the images follow at a time step of 20~s over a period of
8.25~hours.

Another sequence of random fields was prepared from a series of real
(speckle-reconstructed) granulation images obtained with the Dutch
Open Telescope (DOT), La Palma, on 19 October 2001. This series
comprised 198 frames of $14.7\pm 0.1$\,\% rms contrast, measuring
$1040 \times 864$ pixels with a pixel size of 0.071 arcsec and taken
in the 432 nm continuum with a time lag of 30~s between frames. To generate
randomised fields, the images were subject to the following shuffling
procedure. Three series of random numbers ranging between 0.0 and 1.0
were generated by an appropriate standard computer program. The
numbers in two of these series were used to extract randomly-shifted
subfields of $400 \times 400$ pixels from the original field. The
shifts were taken as the above-mentioned random numbers multiplied by
the maximum shifts, 600 pixels in $x$ and 400 in $y$. The third
random number served to compose a new sequence of 198 frames by
randomly shuffling the sequence number of the selected frames, from
which the random subfield was extracted.

In addition to the La Palma granulation images, the random fields
constructed by Rast \cite{rast},
and the sequence of shuffled granulation images, we consider here two
series of artificial granulation patterns obtained by numerical
simulation of granular-scaled convection. The first one was prepared
by Rieutord \etal\ \cite{rieut}. An image of this set is made of
$315\times 315$ pixels (a single pixel being 95.24~km) and covers
a $30\times 30$~Mm$^2$ area. The interval between images corresponds
to 20~s of real time. The second series was computed by M.~Steffen,
B.~Freytag, and H.-G.~Ludwig (see Wedemeyer \etal, \citeyear{stef})
for an $11.2\times 11.2$~Mm$^2$ area; each image is represented by
$200\times 200$ pixels with a pixel size of 56~km and the interval
between frames is 30~s.

\begin{figure}
\centerline{
 \includegraphics[bb=32 2 552 340pt,
 width=0.8\textwidth,clip]{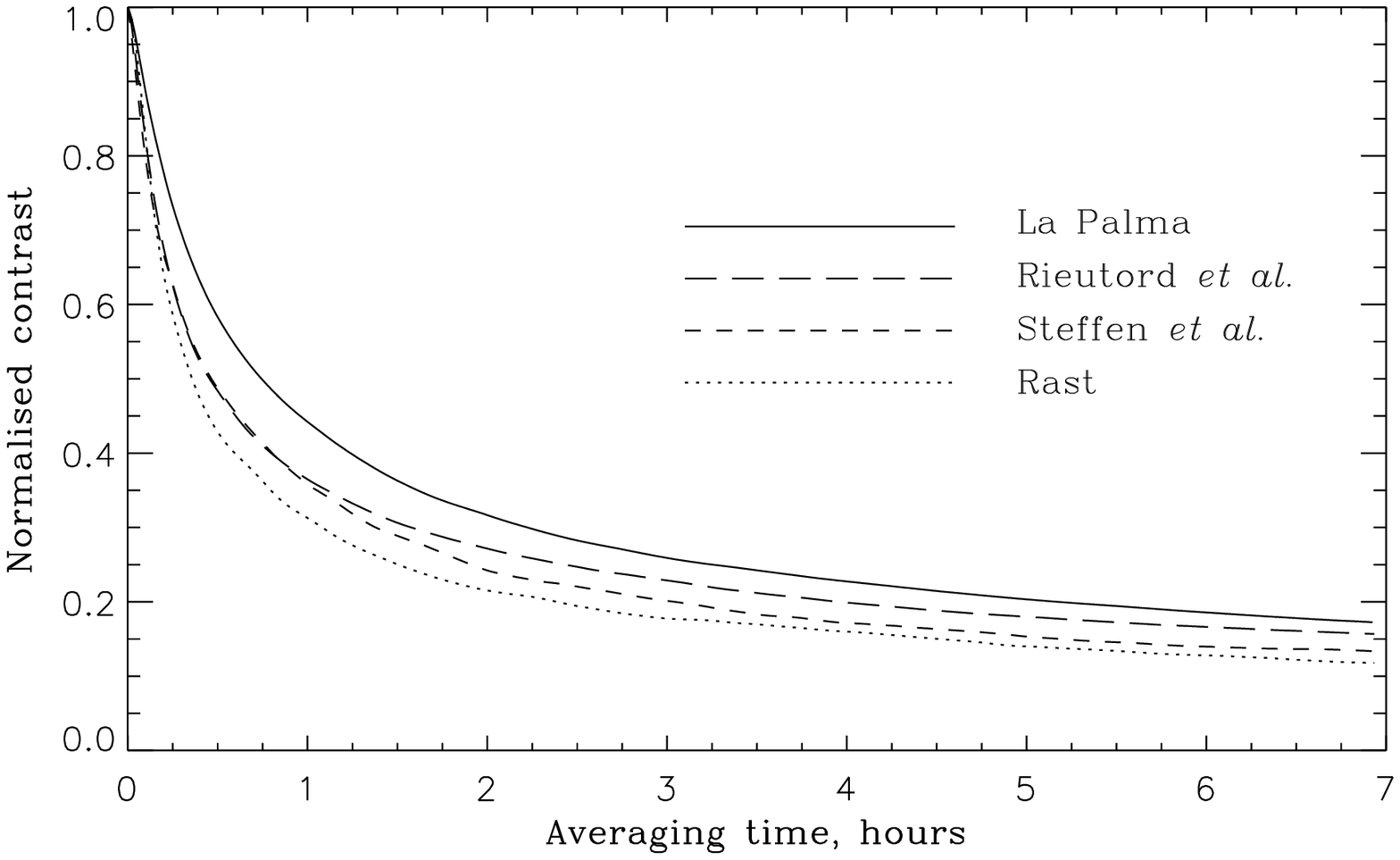}}
 \centerline{\includegraphics[bb=32 6 440 453pt,
 width=0.65\textwidth,clip]{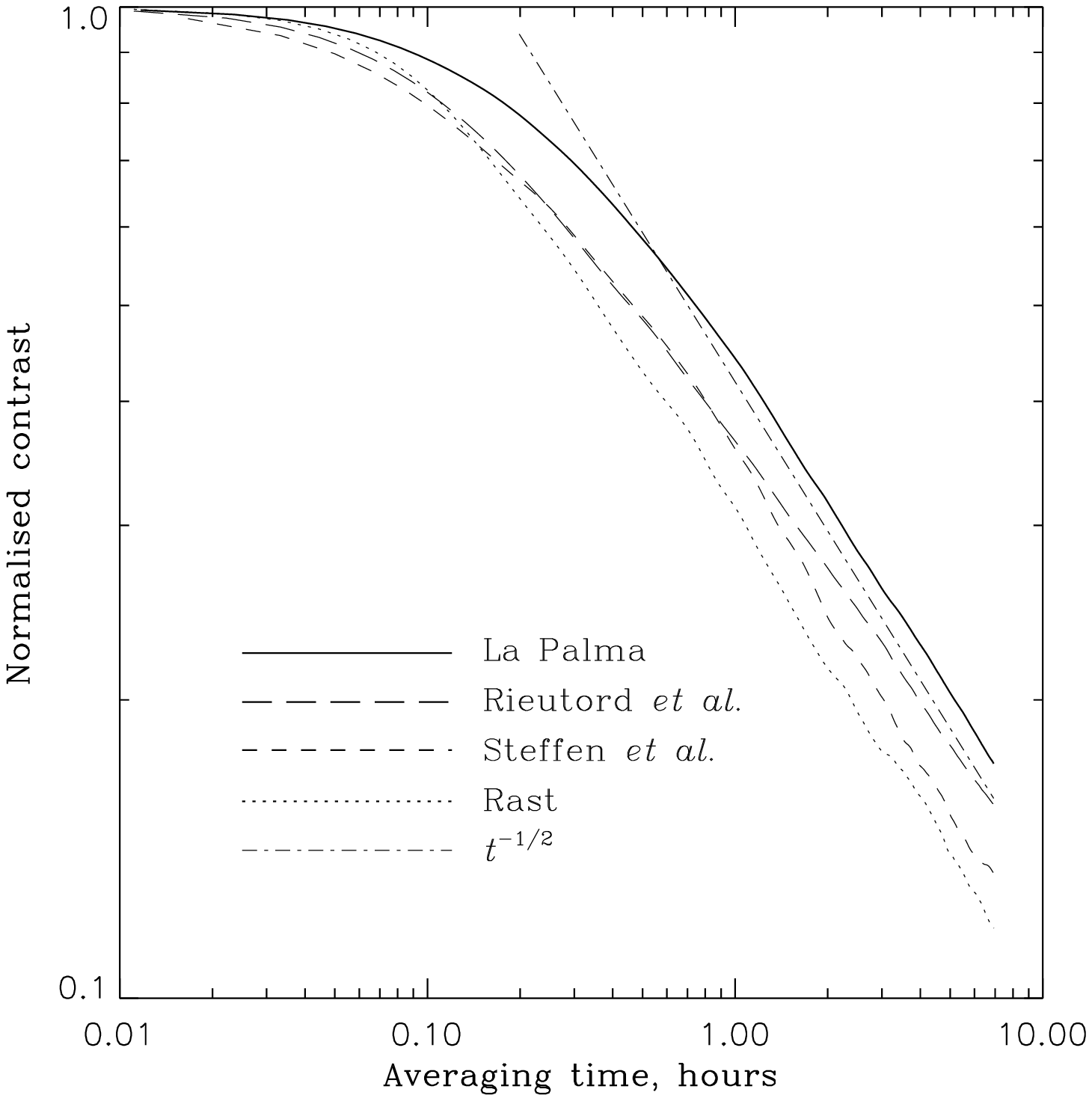}}
   \caption{Contrast of the averaged images as a function of the
   averaging time $t$ for the La~Palma granulation images,
    numerically simulated granulation patterns, and the
    random field constructed by Rast (2002), on a linear
   (\textit{top}) and log--log (\textit{bottom}) scales.
   In the bottom graph, a $t^{-1/2}$ law of variation is also
   shown for comparison.}\label{contr}
\end{figure}

\section{Contrast of Averaged Images}

We note that the rms contrast of the individual images in the La
Palma series varies widely, and even the averaged images exhibit a
large spread in the contrast values, depending on the central time of
the averaging interval. This raises considerable difficulties as the
laws of contrast variation with the averaging time are compared
between the real granulation, the artificial field (Rast
\citeyear{rast}), and simulated
granulation (as noted in the Introduction, it is on such comparisons
that Rast \cite{rast} largely based his criticism on the inferences
of the original paper by Getling and Brandt \cite{gb1}).

For this reason, to obtain unambiguous results, we normalize both the
mean intensity and the rms intensity contrast of each image to a
standard value for either of these two quantities. We apply this
procedure to the four above-mentioned data series (the normalisation
of the DOT series was part of the speckling technique). In addition,
we remove the residual mean-intensity gradient within each frame
(such gradients were noticeable in the images of the La Palma
series).

For the data series so normalised, we compute the variation in the
rms contrast with the averaging time $t$ (at a fixed mid-time of the
averaging interval and with a maximum averaging time of about
7~hours). In the case of the shuffled images, we calculate the
average of $n$ images, from the first to the $n$-th one, the contrast
of this average, and the coefficient of correlation between the first
and the $n$-th image; we represent both the contrast and the
correlation coefficient as a function of $n$.

The resulting contrast-variation curves are plotted in
Figure~\ref{contr}, on both linear (Figure~\ref{contr}a) and log--log
(Figure~\ref{contr}b) scales. The rms contrast of an individual image
is set to unity in all cases; in this sense, the contrast values
presented here can be called normalised. For comparison, in
Figure~\ref{contr}b, we also show a straight line that corresponds to
the statistical $n^{-1/2}$ law of decrease in the rms contrast of $n$
averaged completely random fields with $n$, or, in other words, to
the $t^{-1/2}$ law of contrast decrease with the averaging time $t$.

It can be seen that the curve for the real solar images runs well
above the other three curves. It diverges most widely with the curve
for the random field constructed by Rast \cite{rast}, while the curves for
the numerically
simulated granulation remain in between these extremes. Although the
differences in the slope are most pronounced in the region of
$t\lesssim 1$~h, they remain quite appreciable over the entire range
of averaging times.

\begin{figure}
\centerline{
 \includegraphics[bb=0 0 504 360pt,
 width=0.8\textwidth,clip]{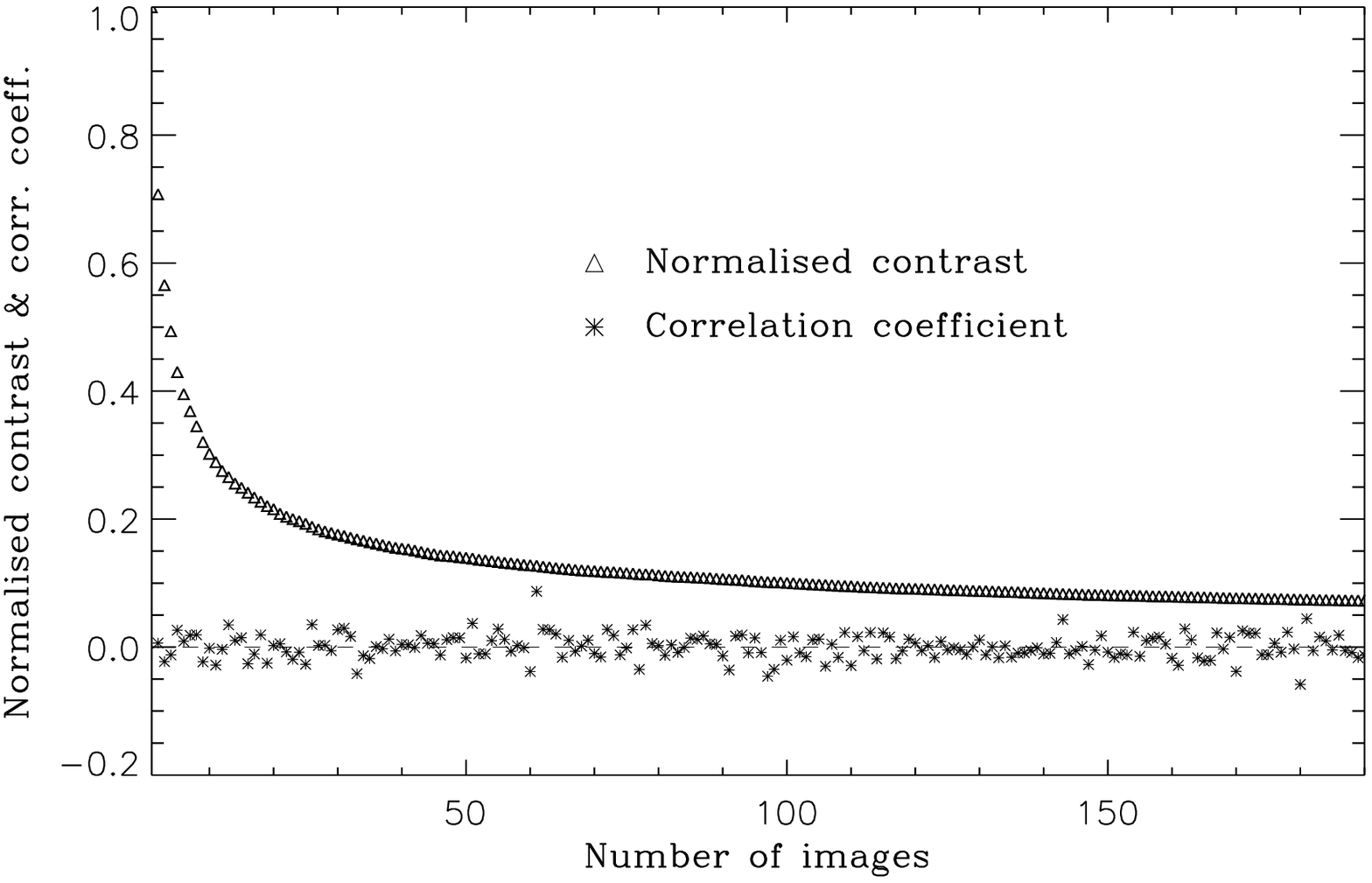}}
   \centerline{\includegraphics[bb=0 0 504 360pt,
 width=0.8\textwidth,clip]{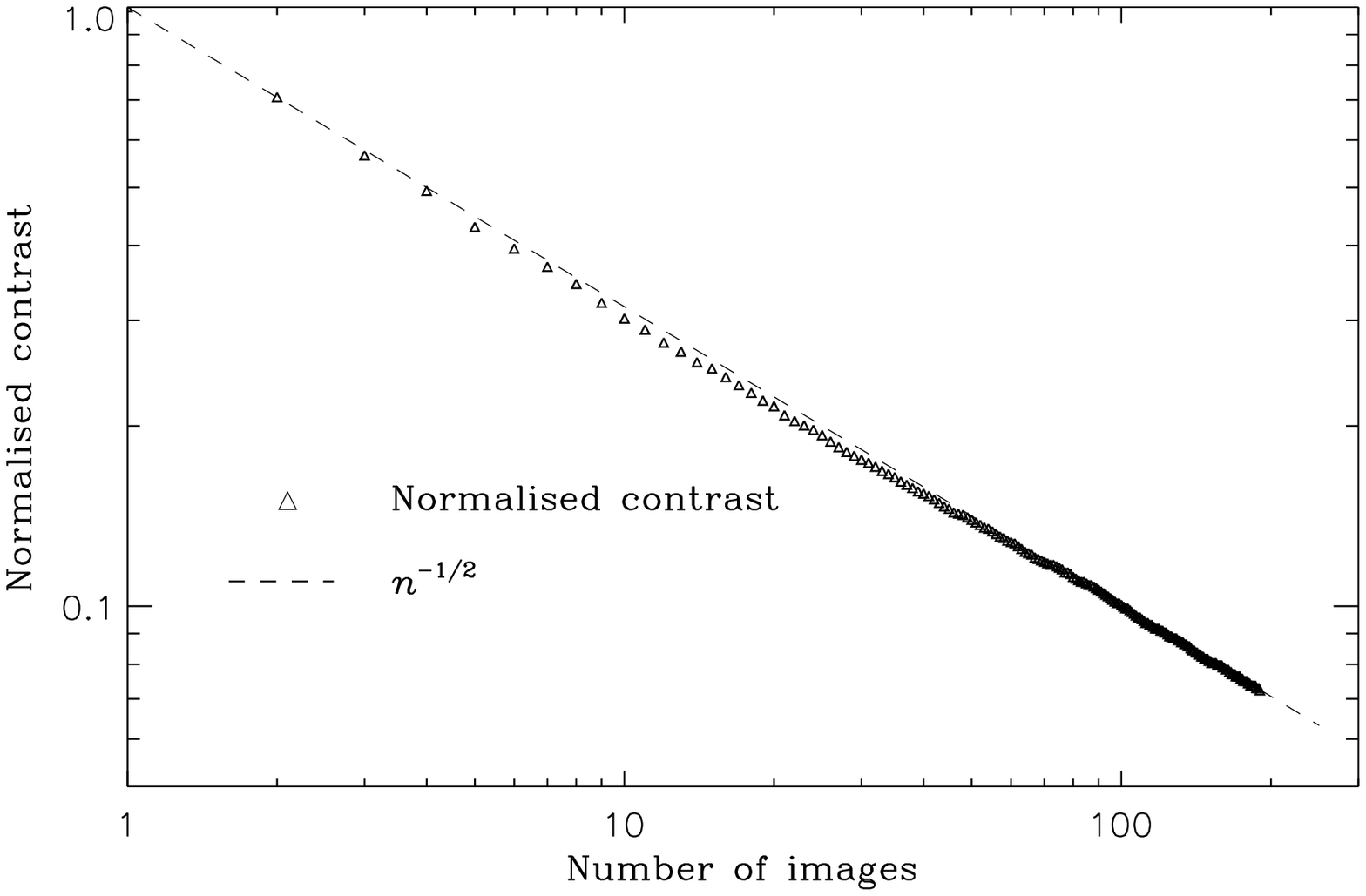}}
    \caption{\textit{Top}: contrast of the averaged sequence of
    $n$ shuffled granulation images and the coefficient of correlation between
    the first and last images in this sequence as a function of
    the number $n$ on a linear scale.
    \textit{Bottom}: same contrast as a function of $n$ but on
    a log--log scale and an $n^{-1/2}$ law of variation.}
    \label{contr_shuffl}
\end{figure}

Most instructive are comparisons of the slopes of the
contrast-variation curves with the slope of the $t^{-1/2}$ straight
line in Figure~\ref{contr}b. The contrast of the real granulation
images decreases with $t$ more slowly than according to the
$t^{-1/2}$ law even in the rightmost part of the graph, at averaging
times of several hours. At the same time, the curve for the field
constructed by Rast \cite{rast}
is nearly parallel to the $t^{-1/2}$ line for $t\gtrsim 0.3$~hours. A
similarly rapid decline can also be noted for one of the series of
simulated images (Wedemeyer \etal\ \citeyear{stef}); it is virtually
consistent with the statistical law in
the region of $t\gtrsim 1$~hour. As for the images obtained by
Rieutord \etal\ \cite{rieut} in their simulations, the corresponding
curve also runs well below that of the La Palma series, although,
among the four curves in Figure~\ref{contr}, it exhibits the slowest
decline at $t\gtrsim 0.8$~hours, even approaching the La Palma curve
at large~$t$.

The fact that the relatively slow decline in the contrast values
should obviously be associated with the presence of some long-lived
component in the granulation field is additionally illustrated by
Figure~\ref{contr_shuffl}, which shows the variation in the rms
contrast of averaged sequences of $n$ shuffled granulation images as
a function of the number $n$. This variation is plotted on a linear
scale in Figure~\ref{contr_shuffl}a and on a log--log scale in
Figure~\ref{contr_shuffl}b. In addition, Figure~\ref{contr_shuffl}a
presents the coefficient of correlation between the first and the
$n$th frame in the sequence as a function of $n$. A straight line
corresponding to the statistical $n^{-1/2}$ law is also plotted for
comparison in Figure~\ref{contr_shuffl}b.

Two remarkable features can be noted immediately. First, the
correlation coefficient of the shuffled images is very small even at
very small $n$ (\eg, $n=2$) and does not exhibit any systematic
trend, fluctuating about zero throughout the entire interval of $n$.
This means that the shuffled images can be considered to be
completely uncorrelated. Second, in agreement with the first
property, they follow very closely the $n^{-1/2}$ over the whole
interval of $n$, again starting from small $n$, such as $n=2$ (the
very small depression of the curve in the range $10\lesssim n
\lesssim 30$ can be attributed to the incomplete validity of the
statistical law at such moderate values of $n$). It is noteworthy
that the field constructed by Rast \cite{rast}, as noted above,
follows the $n^{-1/2}$ law
only from averaging times of about 0.3~hours, while it exhibits a
slower contrast decline at shorter averaging times; obviously, this
effect results from the finite lifetimes of the brightness peaks in
the field constructed by Rast \cite{rast}, which makes the images
within these lifetimes
correlated rather than statistically independent. In contrast, the
shuffling procedure appears to completely destroy any long-lived
features that might be present in the original series.

\section{Discussion and Conclusion}

We see that the contrast of the averaged images of solar granulation
decreases with the averaging time considerably more slowly than does
the contrast of the averages of either a random field similar to the
granulation in some of its parameters (Rast, \citeyear{rast}), or
sequences of shuffled granulation images, or the granulation pattern
numerically simulated by Steffen and his colleagues (see Wedemeyer \etal\
\cite{stef}). This difference is especially pronounced at short to
moderate averaging times.

The artificial granulation pattern obtained in simulations by
Rieutord \etal\ \cite{rieut} differs markedly from the other
computed pattern (Wedemeyer \etal\ \citeyear{stef}) in the
behaviour of the contrast of
averaged images. The former series of images exhibits, at $t\gtrsim
0.8$~hours, the slowest decline in the rms contrast with the
averaging time $t$ (among all the five datasets considered here) and
thus appears to contain a persistent component, while the latter
demonstrates a contrast variation consistent with the statistical
$t^{-1/2}$ law at sufficiently large $t$. This fact remains currently
unexplained. In principle, effects of the sort could be attributed to
an implicit, uncontrolled influence of the boundary conditions used
in the simulations; these conditions can favour the persistence of
some features in the simulated velocity field. However, checking this
conjecture would require a highly scrupulous analysis of the
numerical model. In the context of the adequacy of the so-called
realistic simulations of solar convection, it would be worth doing
such an analysis in the future.

The results of our comparative analysis of the contrast-variation
laws indicate quite definitely that the brightness field of solar
granulation contains a long-lived component. Our approach cannot
distinguish between particular forms of this component. One of us
(P.N.B) adheres to the point of view that it is formed by locally
persistent dark intergranular holes (Roudier \etal,
\citeyear{roudier}; Hoekzema \etal, \citeyear{hbr}; Hoekzema and
Brandt, \citeyear{hb}). The other author of this paper (A.V.G.) is
inclined to attribute the observed contrast-variation law to the
presence of quasi-regular structures (first reported by Getling and
Brandt \cite{gb1} and discussed in detail by Getling in Paper I).
Most likely, it should not be assumed that these two interpretations
contradict each other, since features similar to intergranular holes
can naturally be present in quasi-regular structures.

Our comparison between the contrast-variation curves for the real
granulation and the random field constructed by Rast \cite{rast}
completes the discussion in Paper
I concerning the critical comments by Rast \cite {rast} on our
original paper (Getling and Brandt, \citeyear{gb1}).
Based on all our reasoning, we can dismiss the suggestion by Rast
\cite{rast} on the purely statistical nature of the quasi-regular
structures in the granulation field.


\begin{acknowledgements}

      We are obliged to R.A.~Shine and P.~S\"utterlin for making
      the SOHO MDI and DOT data, respectively, available to us. We
      thank T.~Roudier, M.~Steffen, and M.P.~Rast for putting the data
      of their computations at our disposal. We are also grateful to
      R. Hammer for useful remarks on the text. The work of A.V.G. was
      supported by the \textit{Deut\-sche For\-schungs\-ge\-mein\-schaft}\
      (Project No. 436 RUS 17/56/03) and by the Russian Foundation for Basic
      Research (Project Nos. 04-02-16580-a and 07-02-01094-a).

\end{acknowledgements}

\end{article}
\end{document}